# A New Type of Shape Instability of Hot Nuclei and Prompt Nuclear Fragmentation

J. Tõke and W.U. Schröder

*Department of Chemistry and Nuclear Structure Research Laboratory,*

*University of Rochester, Rochester, New York 14627*

## Abstract

A novel mechanism of prompt nuclear fragmentation is proposed. Assuming microcanonical equilibrium, it is shown that a strong enhancement of the accessible volume of the phase space due to the diffuseness of nuclear surface leads to dynamical instabilities of hot nuclei and to a prompt fragmentation. Equations are derived for the transition temperature $T_T$ for which the thermodynamical surface tension vanishes, as well as for the thermodynamical fissility parameter $\chi_{td}$.

The understanding of properties and the behavior of hot nuclear matter, apart from its general scientific merit, is of key importance in studies of nuclear multifragmentation.[1-3] The latter studies have produced experimental evidence that, under stress generated by heavy-ion collisions, nuclei fragment into multiple pieces - intermediate-mass fragments (IMFs). At the same time, theoretical effort has been undertaken to establish the nature of the stress necessary or sufficient for the loss of (shape) stability of finite nuclei. While theoretical modelling of thermostatic properties of finite nuclear matter[4] has lead to the realization that above a certain critical temperature, $T_{cr}$, nuclear matter cannot exist in its basic liquid phase, most models of nuclear multifragmentation[5-10] rely on the presence of some dynamical stimulus in addition to a purely thermal one. This is so, because the predicted magnitude of $T_{cr}$ appears to be significantly higher than the experimentally determined temperatures of multifragmenting systems. For example, calculations assuming Skyrme interactions predict $T_{cr}$ in the range of 13 – 20 MeV for infinite matter, while the experimentally observed "multifragmentation temperatures" are in the range of 4 – 5 MeV.

The present paper points out the existence of an effect that could lead to the loss of macroscopic stability of finite nuclei at excitation energies of a few MeV per nucleon even in the absence of dynamical (compressional, inertial) stimuli. Its findings derive from a realization of the importance of surface effects in thermodynamics of hot nuclei.

To demonstrate the essence of the new mechanism, a schematic model is adopted in which an excited nuclear system is allowed to assume one of only two macroscopic configurations (phases), that of a spherical mononucleus and that of a dinuclear configuration of two identical touching spheres. It is further assumed that the system is in microcanonical equilibrium, i.e., all micro-states belonging to the allowed macroscopic configurations are populated with equal probabilities. Additionally, to simplify the calculations, it is assumed that the two constituents of the dinuclear configuration have approximately equal excitation energies

$$E_1^* = E_2^* = \frac{1}{2}(E_{tot}^* - E_{pot}), \tag{1}$$

where $E^*_{tot}$ is the total excitation energy of the system and $E_{pot}$ is the potential energy of the dinuclear configuration relative to the ground state of the mononuclear configuration ($E_{pot}=0$). The quantity $E_{pot}$ can be calculated based on the ground-state binding energies of the spherical nuclei involved in both types of configurations and on the Coulomb repulsion energy of the dinuclear complex.

The role of the nuclear surface is described in the present model by the nuclear mass tables, by the liquid-drop mass formula, and by the surface term in the Fermi-gas model expression[11-13] for the level density parameter $a$. It is the latter term that leads to the effects discussed:

$$E = E_V + E_S + E_C(Shape) = \epsilon_V A + \epsilon_S A^{2/3} F_2 + E_C(Shape) \quad and \quad (2)$$

$$a = a_V + a_S = \alpha_V A + \alpha_S A^{2/3} F_2, \quad (3)$$

where A is the atomic number, $E_C(Shape)$ is the shape-dependent Coulomb energy, and $F_2$ is the surface area in units of its value for the spherical shape.

In microcanonical equilibrium, macroscopic states of the system are populated according to weight factors that can be expressed as $W_k \propto e^{S_k}$, where $S_k$ is the entropy of the system in the $k$-th macroscopic configuration. Within the Fermi-gas model, the entropy for the two allowed configurations can be approximated as:

$$S_m = 2\sqrt{a_m E^*} \quad and \quad (4)$$

$$S_d = 4\sqrt{a_d(E^* - E_{pot})/2}, \quad (5)$$

where subscripts $m$ and $d$ identify the mono- and dinuclear configurations, respectively, and the level density parameters are calculated from Eq. 3 for the mass number $A$ (mononuclear) and for the mass number $A/2$ (dinuclear). In Eq. 5, the small contribution to the entropy from the degrees of freedom of relative motion of the constituents of the dinuclear complex has been neglected.

The results of schematic calculations for a hypothetical nucleus with A=200 and Z=80 are shown in Fig. 1. In these calculations, a potential energy of $E_{pot}$=62 MeV was assumed, based on the nuclear mass tables and the Coulomb interaction of two point nuclei of charges $Z/2$, separated by a distance of $d = 2.6(A/2)^{1/3}$ fm. Note that the assumed potential energy is significantly higher than the actual saddle energy for this system. The use of such a high value of $E_{pot}$ in the schematic calculations allows one to better illustrate the large magnitude of the discovered effect. For the level density parameter $a$, the parameterization of Tõke and Swiatecki[11] was employed with $\alpha_V=0.068$ MeV$^{-1}$ and $\alpha_S=0.274$ MeV$^{-1}$.

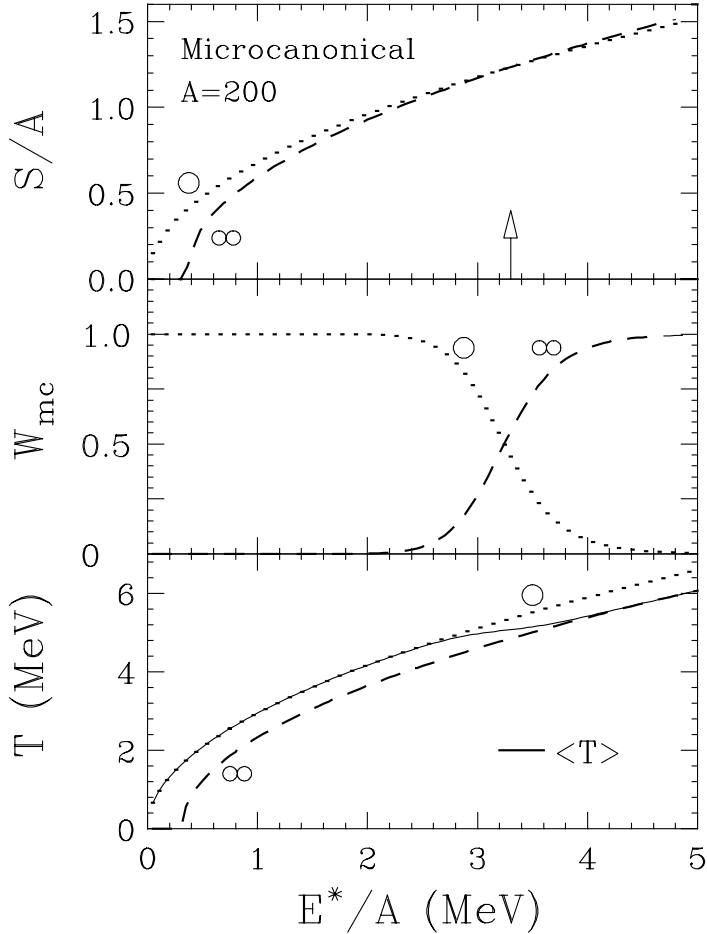

Fig. 1. Entropy per nucleon (top), normalized microcanonical population probability (middle), and temperature (bottom) are plotted vs. total excitation energy per nucleon. Two competing geometries of a nuclear system of A=200, Z=80 are illustrated, a mononuclear (single circle) and a dinuclear (touching circles) configuration. The solid line in the bottom panel represents the average temperature of the system. See text.

As seen in the top panel of Fig. 1, at low total excitation energies, the system achieves the highest entropy when it assumes the mononuclear configuration. In other words, the accessible volume of the phase space is larger for the mononuclear than for the dinuclear configuration. However, the accessible phase space volume is enhanced due to the surface diffuseness of the nuclear matter distribution (reflected in the surface term in Eq. 2). This accessible volume grows faster with increasing total excitation energy for the dinuclear than for the mononuclear configuration. Eventually, at an excitation energy of $E^*/A \approx 3.3$ MeV, the two allowed configurations fill equal phase space volumes, i.e., correspond to equal entropies. Above this "cross-over energy" the system has a higher entropy in the dinuclear state. Obviously, driven by Coulomb repulsion, the latter configuration will decay dynamically, i.e., promptly on the scale on which the microcanonical equilibrium is established.

The middle panel of Fig. 1 illustrates the dependencies of the normalized microcanonical weight factors $W_{mc}$ for the mononuclear and the dinuclear configuration on the total excitation energy. A second-order phase transition from the mononuclear to the dinuclear phase is seen to occur in the smooth, gradual manner characteristic of small systems. This figure demonstrates that the present schematic system cannot survive in a microcanonically equilibrated mononuclear configuration when excited to energies in excess of 4 MeV/nucleon.

The bottom panel of Fig. 1 illustrates the predicted relation (solid line) between average temperature and total excitation energy of the system, i.e, the "caloric curve" for the system. For comparison, the temperatures corresponding to pure mononuclear and dinuclear configurations are also shown. As expected for a microcanonical system, the temperature is not a sharply defined quantity. For any excitation energy, the two-phase system assumes two different temperatures, with probabilities given by the weight functions depicted in the middle panel of Fig. 1. In the caloric curve, the mononuclear-to-dinuclear phase transition shows up as a quasi-plateau around $E^*/A \approx 3.3$ MeV. This quasi-plateau should not be confused with a plateau expected for a first-order phase transition such as, e.g, liquid-to-gas transition.

A very similar behavior is obtained when a canonical, rather than a microcanonical equilibrium is considered for the present system. This is illustrated in Fig. 2, where the free energy per nucleon, the normalized canonical weight factors $W_{can}$, and the total excitation energy per nucleon is plotted vs. temperature for the two configurations, mononuclear and dinuclear.

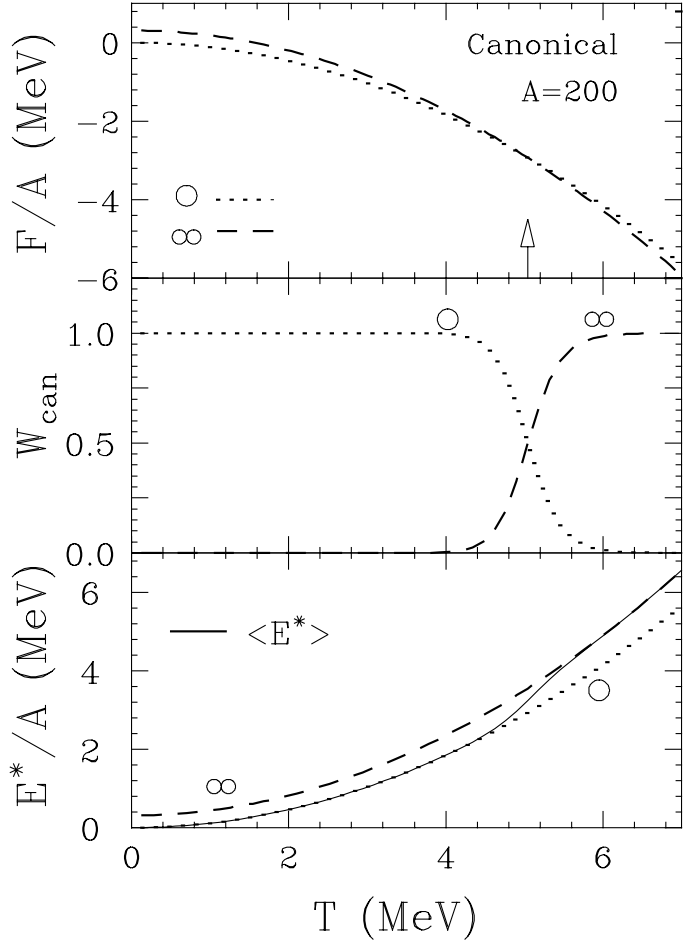

Fig. 2. Free energy per nucleon (top), normalized canonical population probability (middle), and total excitation energy per nucleon (bottom) are plotted vs. temperature. Two competing geometries of a nuclear system of A=200, Z=80 are illustrated, a mononuclear (single circle) and a dinuclear (touching circles) configuration. The solid line in the bottom panel represents the average total excitation energy per nucleon.

In the "canonical" calculations, the free energy was approximated by the non-interacting Fermi-gas model

$$F_m = -a_m T^2 \qquad \text{and} \qquad F_d = -2a_d T^2 + E_{pot} \tag{6}$$

and the weight factors for the mononuclear and dinuclear configurations were assumed to be proportional to $e^{-F_m/T}$ and $e^{-F_d/T}$, respectively.

As seen in the middle panel of Fig. 2, the "canonical" transition from the mononuclear to the dinuclear phase is expected to occur at a temperature of $T \approx 5$ MeV, which is in agreement with the results of microcanonical calculations depicted in Fig. 1. While a microcanonical description appears better suited[6] for isolated nuclear systems than a canonical one, the present schematic model does not reveal any qualitative or major quantitative differences in the behavior of the system in these two approximations.

It is remarkable that already an excitation energy of the order of 4 MeV/nucleon, corresponding to an average temperature of less than 6 MeV, is sufficient for the system to overcome a potential barrier of over 60 MeV. This should not be surprising when one realizes that the mechanism that allows in the present case the system to overcome a large potential barrier is fundamentally the same as that causing thermal expansion of nuclear matter. For example, in a schematic model such as the Expanding Emitting Source Model (EESM),[7] the thermal pressure that causes the system to expand arises as a result of a strong dependence of the level density parameter $a$ on the nuclear matter density $\rho$, $a = a_o(\rho/\rho_o)^{-2/3}$, where $\rho_o$ is the ground state nuclear matter density. In the EESM[7], this thermal pressure is equivalent to potential energies in the compressional degree of freedom of hundreds of MeV, already for temperatures below 10 MeV. Hence, in both cases, a shape-instability (considered in the present schematic model) and a density-instability (considered in the EESM[7]), it is the dependence of the level density parameter $a$ on the "driven" observable (shape and $\rho$, respectively) that generates large effective thermodynamical driving forces and the associated destabilizing potential energies. In both cases, the latter energies are significantly larger than the temperature of the system.

The above $\rho$-dependence of the level density parameter $a$ is not included in the present schematic model, in order to isolate the destabilizing surface effects from other shape-destabilizing effects. In a more complete model, where both, shape and density dependencies of the level density parameter are considered, the loss of shape stability is expected to occur for even lower excitations than indicated in Fig. 1. This is so, because a self-similar[7] radial expansion leads to both, a reduction of the surface energy coefficient $\epsilon_S$ and an enhancement of the surface term in the expression (Eq. 3) for the level density parameter.

To gain a better understanding of the discovered surface effect and its role in generating a shape-instability of finite nuclei, thermodynamical surface tension and thermodynamical fissility are discussed below. The derivation of the respective equations is based on the observation that a thermodynamical driving force $F_\beta$ for a coordinate $\beta$ is generally given by the gradient of the total energy with respect to $\beta$, taken at fixed value of the entropy S. Accordingly, one writes for the thermodynamical surface tension $\Lambda_{td}$

$$\Lambda_{td} = \frac{\partial E^*}{\partial \sigma}\bigg|_{S=const} , \qquad (7)$$

where $\sigma$ is the surface area.

The conditional partial derivative on the right-hand side of Eq. 7 can be calculated by noting that the condition $S = const$ implies

$$\Delta S^2 = 4a^o E^* - 4(a^o + \frac{1}{4\pi r^2}\alpha_S \Delta\sigma)(E^* + \Delta E^* - \frac{1}{4\pi r^2}\epsilon_S \Delta\sigma) = 0 , \qquad (8)$$

where $a^o$ is the ground-state value of the level density parameter, $\epsilon_S$ and $\alpha_S$ are defined via Eqs. 2 and 3, respectively, and $r$ is the radius parameter.

By taking the limit of $\Delta E^*-\,>0$ and $\Delta\sigma-\,>0$, while omitting the terms that are quadratic in these two small quantities, one obtains from Eq. 8

$$\Lambda_{td} = \frac{1}{4\pi r^2}(\epsilon_S - \frac{\alpha_S}{a^o}E^*) = \frac{1}{4\pi r^2}(\epsilon_S - \alpha_S T^2) , \qquad (9)$$

where the Fermi-gas model relationship between temperature $T$ and excitation energy $E^*$, $E^* = a^o T^2$, is utilized.

As seen from Eq. 9, the thermodynamical surface tension $\Lambda_{td}$ decreases monotonically with increasing excitation energy, from its liquid-drop ground-state value of $\Lambda_{ld} = \epsilon_S/4\pi r^2$ to zero at a certain transition temperature $T_T$:

$$T_T = \sqrt{\frac{\epsilon_S}{\alpha_S}} \ . \tag{10}$$

Note that Eq. 10 is analogous to the Fermi-gas model expression for the temperature $T = \sqrt{E^*/a^o}$. A numerical estimate, using $\epsilon_S$=18 MeV and[11] $\alpha$=0.274 MeV$^{-1}$, yields for the transition temperature $T_T \approx 8.1$ MeV, i.e., a value that is significantly lower than (the 13 – 20 MeV) predicted by standard nuclear-matter calculations for semi-infinite matter.[4]

The shape-stability of finite nuclei is commonly described by the fissility parameter $\chi_{ld}$, rather than by the surface tension. The fissility parameter accounts also for the disruptive action of Coulomb forces in addition to the cohesive action of the surface tension. For small ellipsoidal deformations characterized by a shape parameter $\alpha_2$, the surface and Coulomb energies, $E_S$ and $E_C$, are given by

$$E_S = E_S^o(1 + \frac{2}{5}\alpha_2) \ , \qquad E_C = E_C^o(1 - \frac{1}{5}\alpha_2) \ , \tag{11}$$

where $E_S^o$ and $E_C^o$ are the respective energies at a spherical shape. In these terms, the fissility parameter is given by

$$\chi_{ld} = -\frac{\partial E_C}{\partial \alpha_2} \Big/ \frac{\partial E_S}{\partial \alpha_2} = \frac{E_C^o}{2E_S^o} \ . \tag{12}$$

A thermodynamical generalization of Eq. 12 is obtained by replacing the surface energy $E_S^o = 4\pi r^2 \Lambda_{ld} A^{2/3}$ by its thermodynamical counterpart $4\pi r^2 \Lambda_{td} A^{2/3}$:

$$\chi_{td} = \frac{E_C^o}{8\pi r^2 \Lambda_{td} A^{2/3}} = \chi_{ld}(1 - \frac{\alpha_S}{\epsilon_S}T^2)^{-1} \ . \tag{13}$$

Consequently, a spherical nucleus becomes unstable against ellipsoidal distortions when the thermodynamical fissility approaches $\chi_{td} = 1$, i.e., at a limiting temperature of

$$T_{lim} = T_T\sqrt{1 - \chi_{ld}} \ . \tag{14}$$

Here, the quantity $T_T$ is the transition temperature introduced in Eq. 10.

For the present system of $A = 200$, $Z = 80$, Eq. 14 predicts $T_{lim} \approx 4.9$ MeV, when the liquid drop fissility parameter is approximated by $\chi_{ld} = Z^2/50A$. This temperature is consistent with what is seen in the bottom panel of Fig. 1 as necessary or sufficient to cause a transition from the mononuclear to the dinuclear configuration.

In summary, a new surface effect is described that can lead to a loss of macroscopic stability of finite nuclei already at very moderate excitation energies and, hence, to fragmentation. In the constructed schematic microcanonical model of a two-phase system, one observes a second-order phase transition from a mononuclear to a dinuclear configuration. This transition occurs at a temperature that is by more than one order of magnitude lower than the change in potential energy associated with this transition. The large magnitude of the discovered effect calls for a further study of its possible implications. In particular, it appears desirable to include this effect in the practical models of nuclear multifragmentation proposed in the literature. It is worth emphasizing that this surface effect further decreases the stability of hot finite nuclei. It thus amplifies the known effects of a thermal expansion or a reduction of the surface energy coefficient at elevated temperatures.

This work was supported by the U.S. Department of Energy grant No. DE-FG02-88ER40414.